\documentclass[%
 reprint,
superscriptaddress,
 amsmath,amssymb,
prb,
]{revtex4-1}

\usepackage{graphicx}
\usepackage{dcolumn}
\usepackage{bm}

\begin{document}

\preprint{APS/123-QED}

\title{Formation of Cooper pairs between conduction and localized electrons in heavy-fermion superconductors}

\author{Keisuke Masuda}
\email{masuda@kh.phys.waseda.ac.jp}
\affiliation{Department of Physics, Waseda University, Shinjuku, Tokyo 169-8555, Japan}
\author{Daisuke Yamamoto}
\affiliation{Condensed Matter Theory Laboratory, RIKEN, Wako, Saitama 351-0198, Japan}

\date{\today}

\begin{abstract}
Cooper pairing between a conduction electron ($c$ electron) and an $f$ electron, referred to as the ``$c$-$f$ pairing,'' is examined to explain $s$-wave superconductivity in heavy-fermion systems. We first apply the Schrieffer-Wolff transformation to the periodic Anderson model assuming deep $f$ level and strong Coulomb repulsion. The resulting effective Hamiltonian contains direct and spin-exchange interactions between $c$ and $f$ electrons, which are responsible for the formation of the $c$-$f$ Cooper pairs. The mean-field analysis shows that the fully gapped $c$-$f$ pairing phase with anisotropic $s$-wave symmetry appears in a large region of the phase diagram. We also find two different types of exotic $c$-$f$ pairing phases, the Fulde-Ferrell and breached pairing phases. The formation of the $c$-$f$ Cooper pairs is attributed to the fact that the strong Coulomb repulsion makes a quasiparticle $f$ band near the center of the conduction band.

\begin{description}
\item[PACS numbers]
74.70.Tx, 74.20.Mn, 74.25.Dw
\end{description}
\end{abstract}

\pacs{Valid PACS appear here}
\maketitle


\section{\label{sec1}introduction}

Various types of heavy-fermion superconductors discovered recently have attracted growing attention due to their unconventional features. Some materials without inversion symmetry have a superconducting phase in which the mixing of spin-singlet and spin-triplet states is expected.\cite{Bauer,Kimura,Edelstein,Gor'kov} It has also been found that a multilayer material shows a strong-coupling superconducting state where the ratio of the superconducting gap to the transition temperature, $2\Delta/k_{\rm B}T_{\rm c}$, is quite large compared to the conventional BCS value.\cite{Mizukami} Furthermore, possible signatures of the Fulde-Ferrell-Larkin-Ovchinnikov states have been observed in ${\rm CeCoIn_{5}}$.\cite{Y_Matsuda} Although many different heavy-fermion superconductors have been found, the theoretical studies are still insufficient to deeply understand the individual superconducting properties.

To reveal the mechanism of different types of superconductivity, first of all, identifying the pairing symmetry is of crucial importance. In usual heavy-fermion superconductors, the strong Coulomb repulsion between $f$ electrons favors the nodal $d$-wave symmetry, which has been the subject of a number of theoretical studies, including the slave-boson approximation with $1/N$-expansion,\cite{Lavagna,Houghton} the random-phase approximation,\cite{Scalapino,Miyake} the fluctuation-exchange approximation,\cite{Arita} and the third-order perturbation approaches.\cite{Ikeda,Nishikawa,Fukazawa} The experimental results also support the $d$-wave symmetry. The nuclear magnetic and quadrupole resonances (NMR and NQR) in typical heavy-fermion compounds show a power-law temperature dependence of the spin-lattice relaxation rate and the lack of the Hebel-Slichter peak,\cite{Kohori,Mito} which indicate the existence of line nodes. Moreover, the phase diagram has the same feature as that of high-$T_{\rm c}$ cuprates with $d$-wave symmetry; superconductivity appears near the antiferromagnetic phase.\cite{Pfleiderer}

However, conventional $s$-wave superconductivity has also been found in some compounds. In NQR measurements on ${\rm CeRu_{2}}$\cite{K_Matsuda} and ${\rm CeCo_{2}}$,\cite{Ishida_2} the spin-lattice relaxation rate exhibits an exponential decay at low temperatures and shows the Hebel-Slichter peak. Moreover, the recent photoemission spectroscopy (PES) experiment on ${\rm CeRu_{2}}$\cite{Kiss} has shown that the density of states (DOS) has a clear superconducting gap at the Fermi level. All these results were interpreted as evidence for the fully gapped pairing state with $s$-wave symmetry. Usually, this type of simple pairing symmetry can be understood within the framework of the conventional electron-phonon mechanism. However, it is unclear whether the electron-phonon attraction can be dominant since the Coulomb repulsion is rather strong in heavy-fermion systems.

In this paper, we propose another possible way to understand $s$-wave superconductivity in heavy-fermion systems. The essence of our idea is to consider the Cooper pairing between a conduction electron ($c$ electron) and a localized $f$ electron, which we call the ``$c$-$f$ pairing.'' This type of Cooper pairing was previously examined in the study based on a slave-boson approach.\cite{Hanzawa} In this study, since the constraints on the enlarged Hilbert space are treated at the mean-field level, the effects of unphysical states are included in the solution. Using another theoretical treatment, we present a detailed analysis of the $c$-$f$ pairing state, including the derivation of the phase diagram, from a different point of view. By performing the Schrieffer-Wolff transformation to the periodic Anderson model, we first derive an effective Hamiltonian for deep $f$ level and strong Coulomb repulsion. The resulting effective Hamiltonian includes direct and spin-exchange interactions between $c$ and $f$ electrons, which lead to the formation of the $c$-$f$ Cooper pairs. We analyze the effective Hamiltonian within the mean-field approximation and obtain the phase diagrams involving several types of $c$-$f$ superconducting phases. Especially, we find the fully gapped state with anisotropic $s$-wave symmetry in a large region of the phase diagram. We also show that more exotic $c$-$f$ pairing phases, the Fulde-Ferrell (FF) and breached pairing (BP) phases, can appear in the other regions of the phase diagram.

This paper is organized as follows. In Sec. \ref{sec2}, we introduce the periodic Anderson model and derive an effective Hamiltonian by using the Schrieffer-Wolff transformation. We obtain the self-consistent equations for the order parameter of the $c$-$f$ pairing superconductivity and some other quantities within the mean-field approximation. In Sec. \ref{sec3}, we show the results of our numerical calculations. We find three different types of $c$-$f$ pairing phases in the ground-state phase diagram. At the end of the section, we discuss the reason for the formation of those $c$-$f$ pairing states. Finally, Sec. \ref{sec4} is devoted to conclusions.

\section{\label{sec2}model and calculations}
We consider a typical heavy-fermion system composed of itinerant $c$ electrons and nearly localized $f$ electrons, which hybridize with each other. Usually, such a system is modeled by the periodic Anderson Hamiltonian $H_{\rm PAM}=H_{0}+H_{V}$,
\begin{eqnarray}
\nonumber H_{0}&=&-t\sum_{\langle{ij}\rangle}\sum_{\sigma}(c^{\dagger}_{i\sigma}c_{j\sigma}+{\rm H.c.})+\epsilon_{f}\sum_{i\sigma}n^{f}_{i\sigma}\\
&&+U\sum_{i}n^{f}_{i\uparrow}n^{f}_{i\downarrow}-\mu\sum_{i\sigma}(n^{c}_{i\sigma}+n^{f}_{i\sigma}),\label{eq1}\\
H_{V}&=&V\sum_{i\sigma}(f^{\dagger}_{i\sigma}c_{i\sigma}+\rm{H.c.}),\label{eq2}
\end{eqnarray}
where $c^{\dagger}_{i\sigma}$ ($f^{\dagger}_{i\sigma}$) is the creation operator of a $c$ electron (an $f$ electron) with spin $\sigma$ at site $i$, $n^{c}_{i\sigma}=c^{\dagger}_{i\sigma}c_{i\sigma}$, and $n^{f}_{i\sigma}=f^{\dagger}_{i\sigma}f_{i\sigma}$. Here, $t$ is the hopping integral of $c$ electrons, $\epsilon_{f}$ is the position of the bare $f$ level, $\mu$ is the chemical potential, $U$ is the on-site Coulomb repulsion in the $f$ orbital, and $V$ is the hybridization matrix element between $c$ and $f$ states. The sum $\langle{ij}\rangle$ runs over nearest-neighbor pairs of lattice sites. We consider the case of a square lattice in this study.

In order to obtain an effective Hamiltonian describing the $c$-$f$ pairing superconductivity, we perform the Schrieffer-Wolff transformation\cite{Sinjukow} $\bar{H}=e^{S}H_{\rm PAM}e^{-S}$, where $S$ is chosen so as to eliminate all first-order terms in $V$. The generator $S$ must satisfy the condition $\left[S,H_{0}\right]=-H_{V}$, and is given by
\begin{eqnarray}
\nonumber S=&&\frac{1}{\sqrt{\mathstrut N}}\sum_{{\bf k}i\sigma}\left[\frac{Ve^{-i{\bf k}{\cdot}{\bf R}_{i}}}{\epsilon_{{\bf k}}-\epsilon_{f}-U}n^{f}_{i\bar{\sigma}}c^{\dagger}_{{\bf k}\sigma}f_{i\sigma} \right. \\
&&\left. +\frac{Ve^{-i{\bf k}{\cdot}{\bf R}_{i}}}{\epsilon_{{\bf k}}-\epsilon_{f}}\left(1-n^{f}_{i\bar{\sigma}}\right)c^{\dagger}_{{\bf k}\sigma}f_{i\sigma}-{\rm H.c.}\right],\label{eq3}
\end{eqnarray}
where $\bar{\sigma}=\uparrow(\downarrow)$ for $\sigma=\downarrow(\uparrow)$, $N$ is the total number of lattice sites, and $\epsilon_{\bf k}=-2t(\cos{k_{x}}+\cos{k_{y}})$. Here we set the lattice constant $a=1$. When $|\epsilon_{f}|$ and $\epsilon_{f}+U$ are large compared to the effective kinetic energy of $f$ electrons, which is roughly proportional to $\rho{V^2}$ ($\rho$ is the $c$-electron DOS at the Fermi level), the system is approximated by keeping only the zeroth and second orders in $V$ as $\bar{H}{\approx}H_{0}+H_{2}{\equiv}\bar{H}_{\rm eff}$:
\begin{eqnarray}
H_{2}=\frac{1}{2}\left[S,H_{V}\right]=H_{\rm dir}+H_{\rm ex}+H_{\rm ch}+H_{\rm ph},\label{eq4}
\end{eqnarray}
where
\begin{eqnarray}
\nonumber H_{{\rm dir}}&=&\frac{1}{N}\!\sum_{{\bf k'k}i\sigma}
\Bigl(W_{{\bf k'k}}-\frac{1}{4}J_{{\bf k'k}}(n^{f}_{i\uparrow}\!+\!n^{f}_{i\downarrow})\Bigr)\\
&&\;\;\;\;\;\;\;\;\;\;\;\;\;\;\;\;\;\;\;\;\;\;\;\;{\times}e^{-i({\bf k'-k}){\cdot}{\bf R}_{i}}c^{\dagger}_{{\bf k'}\sigma}c_{{\bf k}\sigma},\label{eq5}\\
\nonumber H_{{\rm ex}}&=&\frac{1}{2N}\sum_{{\bf k'k}i}J_{{\bf k'k}}e^{-i({\bf k'-k}){\cdot}{\bf R}_{i}}\Bigl(\!S^{+}_{i}c^{\dagger}_{{\bf k'}\downarrow}c_{{\bf k}\uparrow}\\
&&\;\;\;\;\;\;+S^{-}_{i}c^{\dagger}_{{\bf k'}\uparrow}c_{{\bf k}\downarrow}+S^{z}_{i}(c^{\dagger}_{{\bf k'}\uparrow}c_{{\bf k}\uparrow}\!-\!c^{\dagger}_{{\bf k'}\downarrow}c_{{\bf k}\downarrow})\!\Bigr),\label{eq6}\\
\nonumber H_{{\rm ch}}&=&-\frac{1}{N}\sum_{{\bf k}ij\sigma}
\Bigl(W_{{\bf kk}}-\frac{1}{4}J_{{\bf kk}}(n^{f}_{i\bar{\sigma}}+n^{f}_{j\bar{\sigma}})\Bigr)\\
&&\;\;\;\;\;\;\;\;\;\;\;\;\;\;\;\;\;\;\;\;\;\;\;\;{\times}e^{-i{\bf k}{\cdot}({\bf R}_{i}-{\bf R}_{j})}f^{\dagger}_{j\sigma}f_{i\sigma},\label{eq7}\\
\nonumber H_{{\rm ph}}&=&\frac{1}{4N}\sum_{{\bf k'k}i\sigma}J_{{\bf k'k'}}\\
&&\;\;\;{\times}\Bigl(e^{-i({\bf k'+k}){\cdot}{\bf R}_{i}}
c^{\dagger}_{{\bf k'}\bar{\sigma}}c^{\dagger}_{{\bf k}\sigma}f_{i\sigma}f_{i\bar{\sigma}}+{\rm H.c.}\Bigr).\label{eq8}
\end{eqnarray}
Here, ${\bf S}_{i}=\frac{1}{2}\sum_{\sigma'\sigma}f^{\dagger}_{i\sigma'}{\boldsymbol \tau}_{\sigma'\sigma}f_{i\sigma}$ is the spin operator of $f$ electrons, and the coupling energies $J_{{\bf k'k}}$ and $W_{{\bf k'k}}$ are defined as
\begin{eqnarray}
J_{{\bf k'k}}&=&-\frac{V^{2}}{U}\left(L_{\bf k'}+L_{\bf k}\right),\label{eq9}\\
W_{{\bf k'k}}&=&\frac{V^{2}}{2U}\left(M_{\bf k'}+M_{\bf k}\right),\label{eq10}
\end{eqnarray}
with $L_{\bf k}=U^{2}\left(\epsilon_{{\bf k}}-\epsilon_{f}\right)^{-1}\left(\epsilon_{{\bf k}}-\epsilon_{f}-U\right)^{-1}$ and $M_{\bf k}=U\left(\epsilon_{{\bf k}}-\epsilon_{f}\right)^{-1}$. As shown in Eqs. (\ref{eq4})-(\ref{eq8}), the second term $H_{2}$ consists of four different interaction terms: the direct interaction $H_{\rm dir}$, the spin-exchange interaction $H_{\rm ex}$, the $f$-electron correlated hopping $H_{\rm ch}$, and the pair hopping $H_{\rm ph}$. 

We apply the mean-field approximation to many-body terms in $\bar{H}_{\rm eff}$. We first introduce the following order parameter characterizing $c$-$f$ superconducting phases:
\begin{equation}
\Delta_{\bf k}{\equiv}\frac{1}{N}\sum_{\bf k'}J_{\bf k'k}B_{\bf k'}\label{eq11},
\end{equation}
with
\begin{equation}
B_{\bf k'}={\langle}f^{\dagger}_{{\bf k'+q}\uparrow}c^{\dagger}_{{\bf -k'}\downarrow}-f^{\dagger}_{{\bf -k'+q}\downarrow}c^{\dagger}_{{\bf k'}\uparrow}{\rangle}.\label{eq12}
\end{equation}
By decoupling $H_{\rm dir}$ and $H_{\rm ex}$, one can extract the $c$-$f$ superconducting order parameters, which means that these terms play a crucial role for the formation of the $c$-$f$ Cooper pairs. The effective mass of $c$ electrons is much smaller than that of $f$ electrons. Recently, this type of Cooper pairing with unequal masses has been intensively studied in the field of ultracold Fermi gases.\cite{Liu,Iskin,Parish,Wu,Lin} Liu and Wilczek have discussed this issue by assuming an attractive interaction between fermions with different masses.\cite{Liu} They found that the mean-field phase diagram contains two different types of unconventional superfluid phases as well as the usual fully gapped $s$-wave superfluid phase. One of them is the FF phase with a finite center-of-mass momentum of the Cooper pairs,\cite{Fulde} and the other is the BP phase,\cite{He_2,Gubankova,Forbes} which was called the interior gap superfluid phase in the original paper. In the BP phase, the Cooper pairs have zero center-of-mass momentum, while the Bogoliubov band has no gap unlike the case of the fully gapped $s$-wave state. The name ``breached pairing'' comes from the fact that the superfluid component is ``breached'' by the normal fluid component.\cite{He_2} To take into account the possibility of the FF state, we assume a finite center-of-mass momentum ${\bf q}$ of the Cooper pairs in Eq. (\ref{eq12}). The order parameter $\Delta_{\bf k}$ can be chosen to be real without loss of generality.

In addition to the superconducting order parameter, we also include all possible Hartree-type mean fields, which are defined as
\begin{eqnarray}
\frac{n_{c}}{2}&{\equiv}&{\langle}n^{c}_{i\sigma}{\rangle}=\frac{1}{N}\sum_{\bf k}{\langle}c^{\dagger}_{{\bf k}\sigma}c_{{\bf k}\sigma}{\rangle},\,\,\,\,\,\, \sigma=\uparrow, \downarrow, \label{eq13}\\
\frac{n_{f}}{2}&{\equiv}&{\langle}n^{f}_{i\sigma}{\rangle}=\frac{1}{N}\sum_{\bf k}{\langle}f^{\dagger}_{{\bf k}\sigma}f_{{\bf k}\sigma}{\rangle},\,\,\,\,\,\, \sigma=\uparrow, \downarrow, \label{eq14}\\
\phi^{c}&{\equiv}&\frac{1}{N}\sum_{\bf k}J_{\bf kk}{\langle}c^{\dagger}_{{\bf k}\sigma}c_{{\bf k}\sigma}{\rangle},\,\,\,\,\,\, \sigma=\uparrow, \downarrow, \label{eq15}\\
\phi^{f}&{\equiv}&\frac{1}{N}\sum_{\bf k}J_{\bf kk}{\langle}f^{\dagger}_{{\bf k}\sigma}f_{{\bf k}\sigma}{\rangle},\,\,\,\,\,\, \sigma=\uparrow, \downarrow, \label{eq16}
\end{eqnarray}
where $n{\equiv}n_{c}+n_{f}$ is the total density of the system.
Decoupling the Hubbard term in $H_{0}$ and all the terms in $H_{2}$, we obtain the following mean-field Hamiltonian $\bar{H}_{\rm MF}$:
\begin{eqnarray}
\nonumber \bar{H}_{\rm MF}&=&\sum_{{\bf k}\sigma}\bar{\xi}_{\bf k}c^{\dagger}_{{\bf k}\sigma}c_{{\bf k}\sigma}+\sum_{{\bf k}\sigma}\bar{\epsilon}^{f}_{{\bf k}}f^{\dagger}_{{\bf k}\sigma}f_{{\bf k}\sigma}\\
\nonumber &+&\frac{1}{2}\sum_{\bf k}\Delta_{\bf k}\left(c_{{\bf k}\uparrow}f_{{\bf -k+q}\downarrow}-f^{\dagger}_{{\bf k+q}\uparrow}c^{\dagger}_{{\bf -k}\downarrow}+{\rm H.c.}\right)\\
&-&\frac{1}{4}NUn^{2}_{f}+\frac{1}{2}\sum_{\bf k}\Delta_{\bf k}B_{\bf k}+\frac{1}{2}Nn_{f}\phi,\label{eq17}
\end{eqnarray}
with  $\bar{\xi}_{\bf k}=\epsilon_{\bf k}-\mu+W_{\bf kk}-n_{f}J_{\bf kk}/4$, $\bar{\epsilon}^{f}_{{\bf k}}=\epsilon_{f}-\mu+Un_{f}/2-\phi/2-W_{\bf kk}+n_{f}J_{\bf kk}/4$, and $\phi=\phi^{c}-\phi^{f}$. Note that in Eq. (\ref{eq17}), the effective one-body energy of $f$ electrons, $\bar{\epsilon}^{f}_{{\bf k}}$, depends on the wave vector ${\bf k}$. This means that the correlations between $c$ and $f$ states yield a finite bandwidth for $f$ electrons. From Eq. (\ref{eq17}), we easily find the corresponding thermodynamic potential
\begin{eqnarray}
\nonumber \Omega=&&\sum_{\bf k}\biggl[\bar{\xi}_{\bf k}+\bar{\epsilon}^{f}_{\bf{k}}-\frac{1}{4}Un^{2}_{f}+\frac{1}{2}\Delta_{\bf k}B_{\bf k}+\frac{1}{2}n_{f}\phi\biggr]\\
&&-\frac{2}{\beta}\sum_{{\bf k},\alpha=\pm}\ln{\left(2\cosh{\frac{\beta\mathcal{E}^{\alpha}_{\bf k}}{2}}\right)},\label{eq18}
\end{eqnarray}
where
\begin{eqnarray}
E_{\bf k}&=&\sqrt{(\bar{\xi}_{\bf k}+\bar{\epsilon}^{f}_{{\bf k+q}})^{2}+\Delta^{2}_{\bf k}},\label{eq19}\\
\mathcal{E}^{\pm}_{\bf k}&=&\frac{1}{2}\left(-\bar{\xi}_{\bf k}+\bar{\epsilon}^{f}_{{\bf k+q}}{\pm}E_{\bf k}\right),\label{eq20}
\end{eqnarray}
and $f(E)=1/(e^{{\beta}E}+1)$, with $\beta=1/T$ is the Fermi distribution function. The upper ($\omega^{+}_{\bf k}$) and lower ($\omega^{-}_{\bf k}$) Bogoliubov bands are related to $\mathcal{E}^{\pm}_{\bf k}$ by $\omega^{+}_{\bf k}={\rm max}\left(|\mathcal{E}^{+}_{\bf k}|,|\mathcal{E}^{-}_{\bf k}|\right)$ and $\omega^{-}_{\bf k}={\rm min}\left(|\mathcal{E}^{+}_{\bf k}|,|\mathcal{E}^{-}_{\bf k}|\right)$. The conditions $\frac{\partial \Omega}{\partial B_{\bf k}}=0$, $n=-\frac{1}{N}\frac{\partial \Omega}{\partial \mu}$, $\frac{\partial \Omega}{\partial \phi}=0$, and $\frac{\partial \Omega}{\partial n_{f}}=0$ give the self-consistent equations for $\Delta_{\bf k}$, $\mu$, $n_{f}$, and $\phi$, respectively:
\begin{eqnarray}
\!\!\!\!\!\!\Delta_{\bf k}&=&-\frac{1}{N}\sum_{\bf k'}J_{\bf k'k}\Delta_{\bf k'}\frac{f(\mathcal{E}^{+}_{\bf k'})\!-\!f(\mathcal{E}^{-}_{\bf k'})}{E_{\bf k'}},\label{eq21}\\
\!\!\!\!\!\!n\;&=&2+\frac{2}{N}\sum_{\bf k'}\left(\bar{\xi}_{\bf k'}+\bar{\epsilon}^{f}_{{\bf k'+q}}\right)\frac{f(\mathcal{E}^{+}_{\bf k'})\!-\!f(\mathcal{E}^{-}_{\bf k'})}{E_{\bf k'}},\label{eq22}\\
\!\!\!\!\!\!\nonumber n_{f}&=&\frac{1}{N}\sum_{\bf k'}\Biggl[\left(1+\frac{\bar{\xi}_{\bf k'}+\bar{\epsilon}^{f}_{{\bf k'+q}}}{E_{\bf k'}}\right)f(\mathcal{E}^{+}_{\bf k'})\\
\!\!\!\!\!\!&&\;\;\;\;\;\;\;\;\;+\left(1-\frac{\bar{\xi}_{\bf k'}+\bar{\epsilon}^{f}_{{\bf k'+q}}}{E_{\bf k'}}\right)f(\mathcal{E}^{-}_{\bf k'})\Biggr],\label{eq23}\\
\!\!\!\!\!\!\nonumber \phi\;&=&\frac{1}{N}\sum_{\bf k'}J_{\bf k'}\\
\!\!\!\!\!\!\nonumber &-&\frac{1}{2N}\sum_{\bf k'}\biggl[\left(J_{\bf k'+q}+J_{\bf k'}\right)\left(f(\mathcal{E}^{+}_{\bf k'})\!+\!f(\mathcal{E}^{-}_{\bf k'})\right)\\
\!\!\!\!\!\!\nonumber &&\;\;\;\;\;\;\;\;\;\;\;\;+\frac{\left(J_{\bf k'+q}\!-\!J_{\bf k'}\right)\!\left(\bar{\xi}_{\bf k'}+\bar{\epsilon}^{f}_{{\bf k'+q}}\right)}{E_{\bf k'}}\\
\!\!\!\!\!\!&&\;\;\;\;\;\;\;\;\;\;\;\;{\times}\!\left(f(\mathcal{E}^{+}_{\bf k'})\!-\!f(\mathcal{E}^{-}_{\bf k'})\right)\biggr],\label{eq24}
\end{eqnarray}
with $J_{\bf k}=J_{\bf kk}$. We can see from Eq. (\ref{eq9}) that the order parameter $\Delta_{\bf k}$ can be separated into constant and ${\bf k}$-dependent parts as $\Delta_{\bf k}=\Delta_{0}+\Delta_{1}L_{\bf k}$. Substituting this expression, we derive the equations for $\Delta_{0}$ and $\Delta_{1}$ instead of Eq. (\ref{eq21}):
\begin{eqnarray}
\!\!\!\!\!\!\!\!\!\!\Delta_{0}&=&\frac{V^{2}}{UN}\sum_{\bf k'}L_{\bf k'}(\Delta_{0}+\Delta_{1}L_{\bf k'})\frac{f(\mathcal{E}^{+}_{\bf k'})\!-\!f(\mathcal{E}^{-}_{\bf k'})}{E_{\bf k'}},\label{eq25}\\
\!\!\!\!\!\!\!\!\!\!\Delta_{1}&=&\frac{V^{2}}{UN}\sum_{\bf k'}(\Delta_{0}+\Delta_{1}L_{\bf k'})\frac{f(\mathcal{E}^{+}_{\bf k'})\!-\!f(\mathcal{E}^{-}_{\bf k'})}{E_{\bf k'}}.\label{eq26}
\end{eqnarray}
For a given total density $n$, the values of $\Delta_{0}$, $\Delta_{1}$, $\mu$, $n_{f}$, and $\phi$ are obtained by numerically solving Eqs. (\ref{eq22})-(\ref{eq26}) in a self-consistent way. At the same time, we also need to minimize the free energy $F=\Omega+{\mu}nN$ with respect to ${\bf q}$. In the present study, the center-of-mass momentum ${\bf q}$ is assumed as ${\bf q}=(q/\sqrt{\mathstrut 2},q/\sqrt{\mathstrut 2})$, and the Coulomb repulsion $U$ is set to be $U/t=12$.

\section{\label{sec3}results}
\begin{figure}[h]
\includegraphics[width=8.3cm]{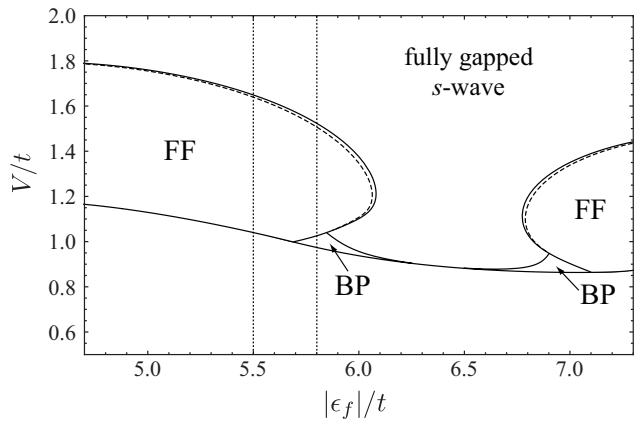}
\caption{\label{ef-V} The phase diagrams in the $(|\epsilon_{f}|/t,V/t)$ plane at $n=2.2$ and $T/t=0.005$. The solid curves indicate the second-order phase transitions between the fully gapped $s$ wave, FF, BP, and normal phases. The dashed curves indicate the position where the gap in the lower Bogoliubov band of the FF phase vanishes. The dotted vertical lines represent the lines of $|\epsilon_{f}|/t=5.5$ and $5.8$, respectively.}
\end{figure}
\begin{figure}[h]
\includegraphics[width=8.0cm]{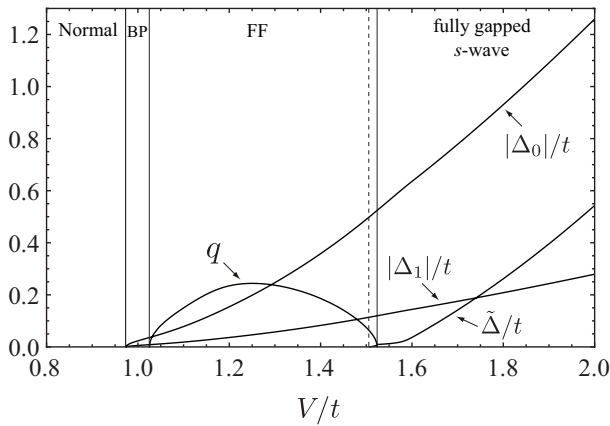}
\caption{\label{q-d0-d1-dt} The $V/t$ dependences of $|\Delta_{0}|/t$, $|\Delta_{1}|/t$, $q$, and $\tilde{\Delta}/t$ at $n=2.2$, $T/t=0.005$, and $|\epsilon_{f}|/t=5.8$. The solid vertical lines indicate the second-order phase transitions. The dashed vertical line represents the position where the gap $\tilde{\Delta}$ of the FF phase vanishes.}
\end{figure}
Before presenting the results of the calculations, we briefly comment on the difference between our study and the previous work by Hanzawa and Yosida.\cite{Hanzawa} Hanzawa and Yosida discussed the $c$-$f$ pairing state on the basis of the periodic Anderson model in the limit of strong Coulomb repulsion, where the doubly occupied states in the $f$ orbital are excluded from the Hilbert space. They derived the gap equation and estimated the order of the transition temperature for the $c$-$f$ pairing superconductivity. In our present study, we further take into account the effect of Hartree-type mean fields and the possibility of the pairing state with finite center-of-mass momentum. Since the Coulomb repulsion is large but finite in our analysis, the influence of doubly occupied states is included in the results. As a consequence, we find several unconventional $c$-$f$ pairing phases in addition to the simple $c$-$f$ pairing phase discussed by Hanzawa and Yosida.
\begin{figure}[b]
\includegraphics[width=7.5cm]{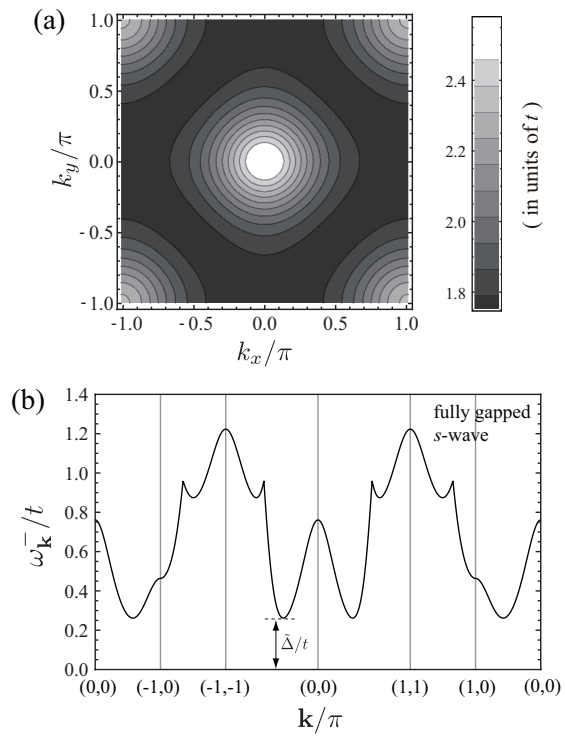}
\caption{\label{full-s} (a) The absolute value of the superconducting order parameter $|\Delta_{\bf k}|$ and (b) the lower Bogoliubov band $\omega^{-}_{\bf k}$ in the fully gapped $s$-wave state at $n=2.2$, $T/t=0.005$, $|\epsilon_{f}|/t=5.8$, and $V/t=1.8$.}
\end{figure}

\begin{figure}[t]
\includegraphics[width=7.0cm]{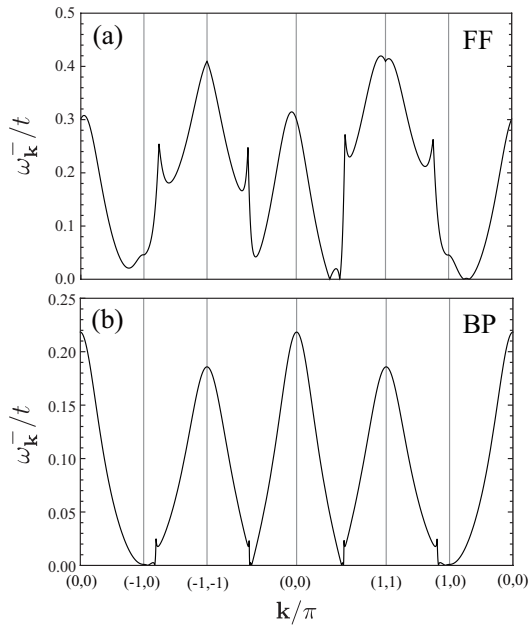}
\caption{\label{ff-bp} The lower Bogoliubov bands at $n=2.2$, $T/t=0.005$, and $|\epsilon_{f}|/t=5.8$ for (a) $V/t=1.3$ and (b) $V/t=1$, respectively.}
\end{figure}

Figure \ref{ef-V} shows the $|\epsilon_{f}|$-$V$ phase diagram at $n=2.2$, which includes four different phases: the fully gapped $s$ wave, FF, BP, and normal phases. Let us discuss the phase transitions between these phases along the line of $|\epsilon_{f}|/t=5.8$, which is depicted by the dotted vertical line in Fig. \ref{ef-V}. We show the $V/t$ dependencies of $\Delta_{0}$, $\Delta_{1}$, and $q$ at $|\epsilon_{f}|/t=5.8$ in Fig. \ref{q-d0-d1-dt}. We also show the actual gap in the lower Bogoliubov band, $\tilde{\Delta}{\equiv}{\min_{\bf k}}\left(\omega^{-}_{\bf k}\right)$. Since the sign of $\Delta_{0}$ is always opposite to that of $\Delta_{1}$ in the parameter range of Fig. \ref{q-d0-d1-dt}, we plotted the absolute values $|\Delta_{0}|$ and $|\Delta_{1}|$ in the figure. For large $V$, the fully gapped $s$-wave phase is preferred. The order parameter $\Delta_{\bf k}$ has anisotropic $s$-wave symmetry and the corresponding lower Bogoliubov band $\omega^{-}_{\bf k}$ shows a finite gap, as shown in Figs. \ref{full-s}(a) and \ref{full-s}(b). As $V$ decreases, the FF state appears as the ground state. Due to the existence of finite $q$, the lower Bogoliubov band $\omega^{-}_{\bf k}$ has an asymmetry with respect to the center of the Brillouin zone, as seen in Fig. \ref{ff-bp}(a). The band is gapless, namely, $\tilde{\Delta}=0$, in most of the FF region. Only in a narrow region ($1.508{\lesssim}V/t{\lesssim}1.524$) of Fig. \ref{q-d0-d1-dt} do we have the fully gapped FF state. As $V$ is decreased further, the transition to the BP phase occurs at $V/t{\approx}1.025$, where the center-of-mass momentum $q$ vanishes. As shown in Fig. \ref{ff-bp}(b), the lower Bogoliubov band $\omega^{-}_{\bf k}$ of this phase touches the zero-energy line, although it is symmetric about the center of the Brillouin zone. For even smaller $V$, we have only the trivial solution $\Delta_{\bf k}=0$, which is natural since the effective attraction $|J_{\bf kk'}|$ between $c$ and $f$ electrons [Eq. (\ref{eq10})] becomes smaller as $V$ is decreased.

It is worthy to note that the $|\epsilon_{f}|$-$V$ phase diagrams for $n=2+\delta$ and $n=2-\delta$ are symmetric with each other about $|\epsilon_{f}|=U/2=6t$. For example, we can obtain the phase diagram for $n=1.8$ by the left-right inversion of Fig. \ref{ef-V}. This symmetric property comes from the fact that the periodic Anderson model has particle-hole symmetry at $\epsilon_{f}=-U/2$ in the case of bipartite lattices.
\begin{figure}[h]
\includegraphics[width=7.5cm]{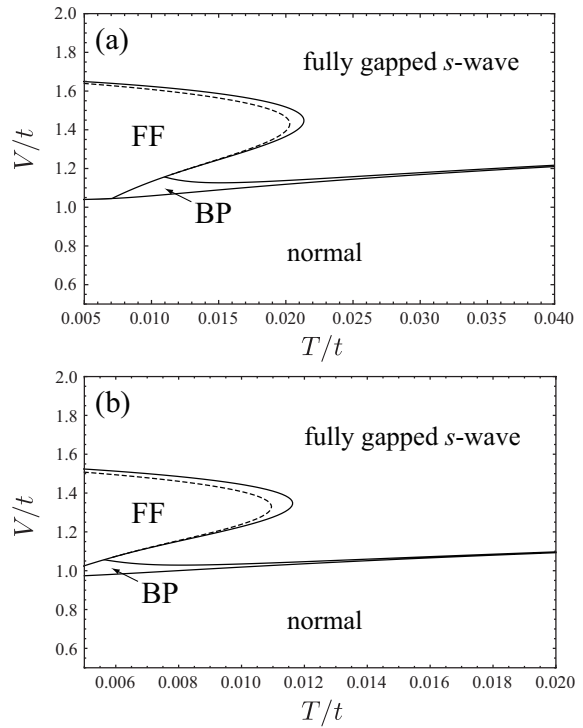}
\caption{\label{T-V} The $T$-$V$ phase diagrams at $n=2.2$ for (a) $|\epsilon_{f}|/t=5.5$ and (b) $|\epsilon_{f}|/t=5.8$. The solid curves indicate the second-order phase transitions between the fully gapped $s$ wave, FF, BP, and normal phases. The dashed curves indicate the position where the gap $\tilde{\Delta}$ of the FF phase vanishes.}
\end{figure}

Next, let us examine the effect of temperature on the $c$-$f$ pairing phases. Figures \ref{T-V}(a) and \ref{T-V}(b) show the phase diagrams in the $(T/t,V/t)$ plane for $|\epsilon_{f}|/t=5.5$ and $|\epsilon_{f}|/t=5.8$, marked by the dotted vertical lines in Fig. \ref{ef-V}. We can see that the fully gapped $s$-wave pairing state is more robust against temperature than the nodal pairing states. Especially, the FF phase completely disappears as the temperature is increased. Such a sensitive temperature dependence of the FF phase has been obtained in previous studies.\cite{Koponen_1,Koponen_2} The region of the BP phase also gets smaller with increasing temperature, but it still survives after the disappearance of the FF phase. For high temperatures, the fully gapped $s$-wave phase occupies a large region of the phase diagram.
\begin{figure}[h]
\includegraphics[width=6.5cm]{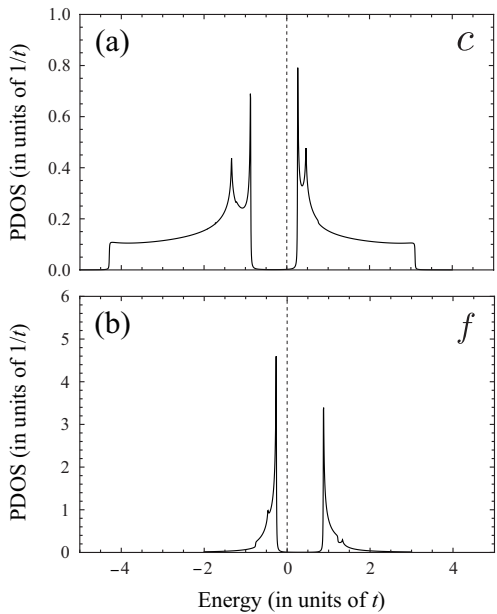}
\caption{\label{pdos} The PDOSs (a) of the $c$ band and (b) of the $f$ band in the fully gapped $s$-wave state at $n=2.2$, $T/t=0.005$, $|\epsilon_{f}|/t=5.8$, and $V/t=1.8$. The dashed vertical lines represent the Fermi level.}
\end{figure}

Since the bare $f$ level $\epsilon_{f}$ is rather deep below the Fermi energy, we should clarify the reason why a $c$ electron near the Fermi level can form a pair with an $f$ electron. We start the discussion from the original periodic Anderson model given by Eqs. (\ref{eq1}) and (\ref{eq2}). Because of the hybridization $V$ between $c$ and $f$ states, the effective $f$ level has a finite dispersion, namely, a finite effective mass, even if the bare $f$ band is completely flat. Furthermore, the strong Coulomb repulsion $U$ splits the effective $f$ level into the upper and lower Hubbard bands and forms a quasiparticle $f$ band in between them, as in the case of the standard Hubbard model.\cite{Georges} The formation of the quasiparticle $f$ band has been shown by the previous studies using the dynamical mean-field theory (DMFT).\cite{Amaricci,Sordi,Medici,Shimizu,Georges} Especially near the half-filling, the quasiparticle $f$ band is generated in the vicinity of the Fermi level, i.e., near the center of the conduction band. This allows us to propose that a conduction electron forms a pair with an electron in the quasiparticle $f$ band and it causes superconductivity.

In the present work, the effect of the Coulomb repulsion $U$ is treated within the mean-field approximation, in which the splitting of the effective $f$ level is not described. However, the quasiparticle $f$ band is approximately expressed by the Hartree shift as $\tilde{\epsilon}_{f}=\epsilon_{f}+Un_{f}/2$. Figures \ref{pdos}(a) and \ref{pdos}(b) show an example of the partial DOSs (PDOSs) of the $c$ and $f$ bands in the fully gapped $s$-wave state. The PDOS of $f$ electrons has a large weight near the Fermi level and the superconducting gap opens in both the PDOSs, which support our scenario proposed above. It should be noted, however, that our mean-field treatment may overestimate the PDOS of the effective $f$ level near the Fermi energy.

It is known that at half-filling the periodic Anderson model has an insulating ground state,\cite{Jarrell,Mutou_1,Pruschke} which exhibits antiferromagnetic order when the Coulomb repulsion is larger than the critical value $U_{\rm c}$.\cite{Rozenberg,Vekic,Horiuchi} In the case of finite doping, the self-consistent second-order perturbation approach by Mutou\cite{Mutou_2} showed that this model favors a metallic ground state, in which the quasiparticle $f$ band is located around the Fermi level. However, this study did not take into account the $c$-$f$ pairing state. We expect that the $c$-$f$ pairing state can appear in a doped region of the periodic Anderson model. In order to discuss the doping-induced phase transition from insulator to $c$-$f$ pairing state, it is required to perform a further analysis which can treat the insulating states, e.g., the use of the DMFT, although it is beyond the scope of the present study.

\section{\label{sec4}conclusion}
We have studied the Cooper pairing between a conduction electron ($c$ electron) and an $f$ electron, called the ``$c$-$f$ pairing,''\cite{Hanzawa} to understand $s$-wave superconductivity in heavy-fermion systems. Considering a system with deep $f$ level and strong Coulomb repulsion, we first derived an effective Hamiltonian by performing the Schrieffer-Wolff transformation to the periodic Anderson model. Within the mean-field analysis of the effective Hamiltonian, we obtained the ground-state phase diagrams including three different types of $c$-$f$ pairing phases: the fully gapped, FF, and BP phases. Especially, we found that the fully gapped $c$-$f$ pairing phase with anisotropic $s$-wave symmetry occupies a large region of the phase diagram. Moreover, we demonstrated that the fully gapped $c$-$f$ pairing state is more robust against temperature than the FF and BP phases. Our results may be relevant to the recent experiment which observed an anisotropic $s$-wave superconducting gap in ${\rm CeRu_{2}}$.\cite{Kiss}

\begin{acknowledgments}
We thank Grant-in-Aid from JSPS (K. M.) and KAKENHI (23840054) from JSPS (D. Y.).
\end{acknowledgments}


\end{document}